\documentclass{aa} 
\usepackage{graphics}
\begin{document}

\def\hi {H\,{\sc i}}
\def\oiii {O{\sc iii}}
\def\fri {FR\,I}
\def\frii {FR\,II}
\def\txs {TXS\,2226{\tt -}184}
\def\pks {PKS\,2322{\tt -}123}
\def\radm {rad m$^{-2}$}
\def\ab {$\sim$}
\def\etal {{\sl et~al.\ }}

\def\dg{$^{\circ}$}
\def\kms{km\,s$^{-1}$}
\def\solmass {\hbox{M$_{\odot}$}}
\def\solum {\hbox{L$_{\odot}$}}

\title{Discovery of a luminous water megamaser in the \frii\ radiogalaxy 3C\,403}
\author{A.\ Tarchi\inst{1,2}
        \and
        C.\ Henkel\inst{3}
        \and
        M.\ Chiaberge\inst{1,4}\thanks{ESA fellow}
        \and K.\ M.\ Menten\inst{3}
        }

\offprints{A. Tarchi,
\email{a.tarchi@ira.cnr.it}}

\institute{Istituto di Radioastronomia, CNR, Via Gobetti 101, 40129 Bologna, Italy
\and Osservatorio Astronomico di Cagliari, Loc. Poggio dei Pini, Strada 54, 09012 Capoterra (CA), Italy
\and Max-Planck-Institut f{\"u}r Radioastronomie, Auf dem H{\"u}gel 69, D-53121 Bonn, Germany
\and Space Telescope Science Institute, 3700 San Martin Drive, Baltimore, MD 21218, USA
}

\date{Received date / Accepted date}

\abstract{We report the first detection of a water megamaser in a radio-loud galaxy, 3C\,403. 
This object has been observed as part of a small sample of \frii s with evidence of nuclear 
obscuration. The isotropic luminosity of the maser is $\sim$1200\,\solum. With a recessional 
velocity of c$z$\,$\sim$\,17680\,\kms\ it is the most distant H$_{2}$O maser so far reported. 
The line arises from the densest interstellar gas component ever observed in a radio-loud galaxy. 
Two spectral features are observed, likely bracketing the systemic velocity of the galaxy. These
may either arise from the tangentially seen parts of a nuclear accretion disk or from dense warm 
gas interacting with the radio jets.
\keywords{Galaxies: individual: 3C\,403 -- Galaxies: active -- 
          Galaxies: ISM -- masers -- Radio lines: ISM -- Radio lines: galaxies}}

\titlerunning{Discovery of a water megamaser in an \frii\ galaxy}

\authorrunning{Tarchi et al.}

\maketitle

\section{Introduction}
So far, water megamasers have been detected in radio-quiet active galactic nuclei (AGN; for a definition of radio-quiet AGN, see e.g.\ Kellerman et al.\ \cite{kellerman89}), mostly in Seyfert\,2 and LINER galaxies. Only one H$_{2}$O 
megamaser may have been found in a Seyfert\,1 type AGN (Nagar et al.\ \cite{nagar02}) and another 
one is known to be associated with a radio-quiet elliptical galaxy, NGC\,1052. Assuming that all 
megamasers are associated with molecular material orbiting around the central engine (e.g.\ Miyoshi 
et al.\ \cite{miyoshi95}) or interacting with the nuclear jet(s) of the host galaxy (see Claussen 
et al.\ \cite{claussen98}) and that their amplification is unsaturated (i.e. the maser intensity 
grows linearly with the background radio continuum), one should expect a much higher detection rate 
in radio-loud AGN than is actually observed.
 
The first systematic search for H$_2$O maser emission from radio-loud galaxies was performed by Henkel 
et al.\ (\cite{henkel98}) in $\sim$50 \fri \footnote{Radio galaxies are classified according to their 
radio morphology: \fri s are edge-darkened sources, while \frii s are edge-brightened (Fanaroff \& Riley 
\cite{fanaroff74}).} galaxies with redshift $z$ $<$ 0.15 but no masers were detected. The most plausible 
and simple explanation is that \fri\ galaxies (or at least the majority of them) lack a geometrically 
and optically thick molecular torus. This scenario is now supported by an increasing number of studies 
(e.g. Chiaberge et al.\ \cite{chiaberge99}, Perlman et al.\ \cite{perlman01}, Whysong \& Antonucci 
\cite{whysong01}, Verdoes Kleijn et al.\ \cite{verdoes02}). 

Within the framework of the AGN unification scheme (e.g. Urry and Padovani \cite{urry95}), narrow-lined 
(NL) \frii s are believed to represent the `parent population' of  radio-loud quasars and broad-lined 
(BL) radio galaxies. In order to account for unification between these objects, the presence of a 
geometrically and optically thick obscuring structure has been proposed, in analogy to the radio-quiet 
unifying picture (Barthel \cite{barthel89}). Samples of radio galaxies belonging to the BL and NL \frii\ 
class have been recently observed with the Effelsberg telescope (Tarchi et al.; Lara et al., priv. comm.). 
As in the case of \fri s, no maser detections have been obtained.

\begin{table*}[ht]
\begin{center}
\caption[]{Target galaxies: Source name, coordinates, redshift, 178\,MHz total flux, 5\,GHz core flux, 
[\oiii] line Equivalent Width, maser luminosity and observed velocity range are given. The data were taken 
from the NASA/IPAC extragalactic database (NED), from Table~1 of Chiaberge et al.\ (\cite{chiaberge02}) 
and references therein, and from the observations presented here.
\label{sources}}
\begin{tabular}{lccccccccc}
\hline
\\
Source       & RA          & Dec          & $z$ & $S_{\rm  178\,MHz}^{\rm tot}$ & ${\rm log}(S^{\rm core}_{\rm 5\,GHz})$      
             & ${\rm log}(EW_{\rm {[\oiii]}})$ & $L$$_{\rm maser}^{\rm a)}$ & $V_{min}^{\rm b)}$ & $V_{max}^{\rm b)}$\\
             & (J2000)     & (J2000)      &   & (Jy) & $\rm (erg\,cm^{-2}\,s^{-1}\,Hz^{-1})$ & ($\rm {\AA}$) & (\solum) 
             & (\kms) & (\kms) \\
\\
\hline
\\
3C\,98    & 03 58 54.4 & +10 26 02    & 0.0304    & 35.5 & $-$24.99       & 4.463  & $<$ 2.8    & 8700 & 9800\\
3C\,285   & 13 21 17.8 & +42 35 15    & 0.0794    & 6.0 & $-$25.11        & 4.385  & $<$ 6.3    & 23300 & 24400\\
3C\,403   & 19 52 15.8 & +02 30 24    & 0.0590    & 17.8 & $-$24.94       & 4.181  & $\sim$1200 & 17000 & 18700 \\
\\
\hline
\end{tabular}
\end{center}

a) Maser luminosities assuming isotropic emission and using [$L_{\rm H_2O}$/L$_{\odot}$] = 0.023 $\times$ 
[$\int{S\,{\rm d}V}$/Jy\,km\,s$^{-1}$] $\times$ [$D$/Mpc]$^{2}$. 3$\sigma$ upper limits for for the non-detected 
sources are calculated assuming emission in a single 1.1\,\kms\ wide channel.\\
b) Observed velocity range (optical convention). $V$ is related to the observed frequency by $\nu_{\rm obs} = 
\nu_{\rm rest}\,(1 + \frac{V}{\rm c})^{-1}$.
\end{table*}

In this Letter we report the first detection of water maser emission in a powerful radio galaxy. The H$_{2}$O 
megamaser has been detected in the \frii\ galaxy 3C\,403, observed as part of a small sample of sources described 
in Sect.\,2.
\section{Sample selection}
We have considered  nearby ($z<0.1$) \frii s from the 3CR catalog (Spinrad \cite{spinrad85}), for  which information  
on the radio and optical nuclear emission, and the flux of the [\oiii] emission line are available.

Our sample (Table\,\ref{sources}) comprises all 3 \frii s spectrally classified as High Excitation Galaxies  
(HEGs, Jackson  \&  Rawlings \cite{jackson97}) with nuclear equivalent widths of the [\oiii]$\lambda$5007 emission 
line\footnote{The nuclear EW of the [\oiii]$\lambda$5007 emission line is measured with respect to the nuclear unresolved  
continuum source observed in HST images (Chiaberge et al.\ \cite{chiaberge02})} EW([\oiii])$>10^{4}$ \AA. A high 
value for the nuclear EW([\oiii]) in HEGs has been interpreted as a hint for the obscuration of the central ionizing 
continuum source (Chiaberge et al.\ \cite{chiaberge02}). In these sources the nuclear ionizing continuum would be 
obscured to our line-of-sight and only a small fraction of the emission is seen through scattered light. The same 
indicator has been used to identify absorbed sources among Seyfert galaxies (e.g. Kinney et al.\ \cite{kinney91}). 
Therefore, our selection criteria provide us with a sample of galaxies with both high radio flux densities and 
nuclear obscuration of the central ionizing source.
\section{Observations and data reduction}
Observations of the $6_{16} - 5_{23}$ transition of H$_2$O (rest frequency: 22.23508\,GHz) were carried out with 
the 100-m telescope of the MPIfR at Effelsberg\footnote{The 100-m telescope at Effelsberg is operated by the 
Max-Planck-Institut f{\"u}r Radioastronomie (MPIfR) on behalf of the Max-Planck-Gesellschaft (MPG).} in January 
and March 2003. The beam width was 40$''$. The observations were made in a dual beam switching mode with a beam 
throw of 2\arcmin\ and a switching frequency of $\sim$1\,Hz. The system temperature, including atmospheric 
contributions, was $\sim$20--40\,K on an antenna temperature scale ($T_{\rm A}^*$). The beam efficiency was 
$\eta_{\rm b}$$\sim$0.3. Flux calibration was obtained by measuring W3(OH) (see Mauersberger et al.\ \cite{mauer88}). 
Gain variations of the telescope as a function of elevation were taken into account (Eq.\,1 of Gallimore et al.\ 
\cite{gallimore01}). The pointing accuracy was better than 10\arcsec. All data were reduced using standard procedures 
in the GILDAS software package (http://www.iram.fr/IRAMFR/GS/gildas.htm).
\section{Results}
Table~\ref{sources} summarizes characteristic properties of the three observed sources. Coordinates, redshifts and 
total 178\,MHz fluxes (Cols.\,2$-$5) were taken from the NASA/IPAC Extragalactic Database (NED), while the values 
for the 5\,GHz core luminosity and [\oiii] equivalent width (Cols.\,6$-$7) are taken from Chiaberge et al.\ (2002) and 
references therein. Upper limits for the non-detected sources as well as the isotropic luminosity of the maser 
detected in 3C\,403 are given in Col.\,8. Cols.\,9 and 10 show the velocity ranges of our spectra.

\begin{figure}[ht]
\centering
\resizebox{8.0cm}{!}{\includegraphics{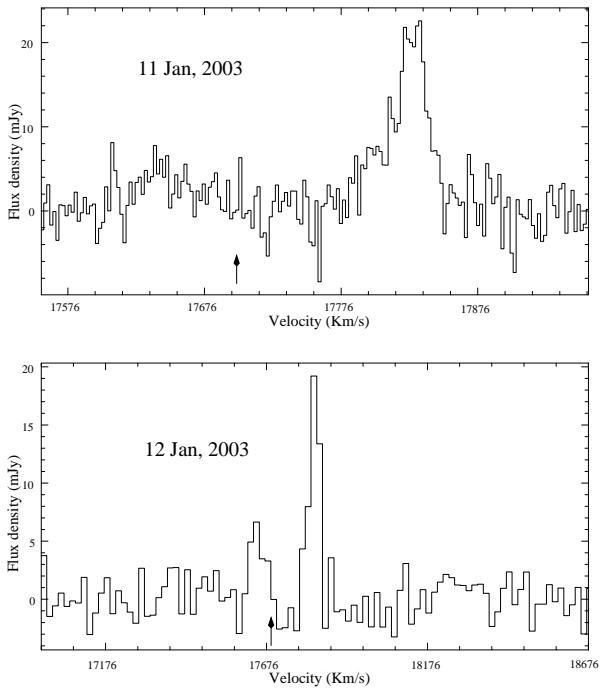}}
\caption{Maser lines in 3C~403 with a channel spacing of 78\,kHz or 1.15\,\kms ({\it upper panel}) and 1.25\,MHz or
17.8\,\kms ({\it lower panel}). The arrow marks the systemic velocity of the galaxy, $V_{\rm sys}$ = c$z$ = 17688 \kms. 
\label{2spec}}
\end{figure}

The maser spectra of 3C\,403, taken in January, are shown in Fig.\,\ref{2spec}. The profile is composed of two main
components (asymmetrically) bracketing the nominal systemic velocity of the galaxy (17688\,\kms, see Sect.\,5.2): the 
stronger one has a velocity of 17827$\pm$1\,\kms, a width of 31$\pm$2\,\kms, and a flux density peak of 23$\pm$3\,mJy; 
the weaker one has a velocity of 17644$\pm$5\,\kms, a width of 53$\pm$8\,\kms, and a peak flux density of 4.0$\pm$0.5\,mJy.
Subcomponents, like the blueshifted shoulder of the main component ($\sim$17790\,\kms) and the feature near 
17610\,\kms\ are also apparent. Using a distance of 235\,Mpc ($H_{\rm 0}$ = 75\,\kms\,Mpc$^{-1}$), the total isotropic 
luminosities are 950$\pm$140 and 280$\pm$55\,\solum\ for the main components, respectively. 

A further observation was performed on March 25. No significant change in the line profile was found.
\section{Discussion}

The standard unified scheme of AGN requires an obscuring region, possibly containing molecular gas that surrounds 
the central engine and that effectively shields the inner few parsecs from view, if the radio axis lies close to 
the plane of the sky (Antonucci \cite{antonucci93}). In the innermost part, at radii up to some tenths of a parsec, 
this material is likely to form a rapidly rotating accretion disk around a central supermassive black hole. At 
larger distances (up to about 50-100 pc) the atomic and molecular gas is possibly distributed in a toroidal structure 
providing obscuration of the central regions to particular lines-of-sight.

As mentioned in Sect.\,1, interferometric studies of H$_2$O megamasers have shown that the emission is either associated 
with a nuclear accretion disk (for NGC\,4258, see e.g.\ Miyoshi et al.\ \cite{miyoshi95}) or with the radio-jets interacting 
with dense molecular material near the center (for Mrk\,348, see Peck et al.\ \cite{peck03}). According to this picture, 
we expect to detect water megamaser activity in \frii\ radio galaxies, where accretion disks are likely present and 
powerful radio jets have been extensively mapped.

The absence of water maser detections among early-type radio galaxies has been discussed by Henkel et al.\ \cite{henkel98}. 
Although this work was focused on \fri\ galaxies, the lack of similar studies on \frii s and the presumed geometrical 
similarity between \fri s and \frii s make their conclusions also relevant for our study. Among the different scenarios, 
Henkel et al.\ proposed that radio galaxies may be lacking molecular gas in the nuclear region. Indeed, very few direct 
detections of molecular gas in radio-loud galaxies have been reported so far ($\rm H_{2}$ in Cygnus\,A: Evans et al.\ 
\cite{evans99}; CO in 3C\,293: Wilman et al.\ \cite{wilman00}). The detection discussed here strongly favors the presence 
of molecular material near the central engines of at least some \frii s. Past negative results in molecular line surveys 
may also be a consequence of observational sensitivity limits (such a possibility was also mentioned by Henkel et al.\ 
\cite{henkel98}). Furthermore, because of the high gas densities required for H$_{2}$O masers to operate ($>$10$^{7}$\,cm$^{-3}$; 
e.g.\ Elitzur et al.\ \cite{elitzur89}), our detection represents the densest interstellar gas component ever observed 
in a radio-loud galaxy (typical values for molecular gas densities range between $\sim 10^{3}$ and $ \sim 10^{5}$ cm$^{-3}$; 
e.g.\ Wilman et al.\ \cite{wilman00}). 

Observational sensitivity limits could also explain the result reported by Morganti et al.\ (\cite{morganti01}) on 3C\,403. 
They searched for neutral hydrogen absorption in a small sample of \fri\ and \frii\ galaxies using the VLA. \hi\ absorption 
was mostly found in NL \frii s, while none was detected in broad-lined (BL) \frii s and only one detection was obtained in 
an \fri. This result is, to first order, consistent with the unified schemes. Interestingly, among the NL \frii s the only 
\hi\ non-detection was 3C\,403 (with a 3$\sigma$ {H{\sc i}} column density upper limit of $\sim$ 5$\times 10^{20}$ $\rm cm^{-2}$, 
using $T_{\rm spin}$=100\,K and a linewidth of 30\,\kms, that is smaller than the linewidths of the associated gaseous
nebulae as e.g.\ observed by Baum et al.\ \cite{baum90}). This non-detection may either be due to the rather high upper 
column density limit or to an intrinsic lack of dense gas in the galaxy's nuclear region. Our maser detection favors
the former possibility.
\subsection{Is 3C\,403 a peculiar \frii?}
While optical photometric and morphological profiles of the galaxy 3C 403 indicate a normal elliptical structure (Govoni et 
al.\ \cite{govoni00}), its radio morphology is quite peculiar. An 8.4 GHz VLA map of 3C\,403 (Black et al.\ \cite{black92} 
(BBL); their Fig.\,13) shows that the radio galaxy has a remarkable X-shape, with `wings' of $\sim$ 100 kpc extending in NW 
and SE directions. A weak jet feature (labeled as F7\&8 by BBL) is present between the core and the exceptionally bright 
knot F6. The lobes and their hot-spots (F1\&2 and P1\&2 in BBL), well displaced from the wings, are also prominent and confirm 
the \frii\ classification. Can 3C\,403 be considered to be a typical \frii? Or does it belong to a peculiar class of objects 
where water masers are more likely found. We also note that 3C\,403 has a larger total radio luminosity (by factors of almost
2) than the other sources in our sample (see Table 1). To discriminate between effects related to radio power or 
morphological peculiarities, more extended H$_{2}$O maser surveys are needed. 
\subsection{Accretion disk or jet interaction?}
For the following discussion it is worth to emphasize the uncertainty of the systemic velocity of 3C\,403. $V_{\rm sys}$
derived from optical emission lines may be uncertain or biased by motions of the emitting gas (e.g. Morganti et al.\
\cite{morganti01}). In view of the rotation curve measured by Baum et al. (\cite{baum90}), however, uncertainties may 
be $<$100\,\kms\ in the case of 3C\,403, so that we assume in the following that the systemic velocity is placed (as 
indicated by Fig.\,\ref{2spec}) in between the two main H$_2$O maser components. Only interferometric observations 
will allow us to pinpoint the location of the H$_2$O emitting region(s), to determine (or to provide an upper limit to) 
the extent of the emission, and to associate the maser with an accretion disk or the radio jets. Nevertheless, a 
qualitative discussion is possible on the basis of the single-dish spectra of Fig.~\ref{2spec}.  

The handful of bright knots visible in the radio image shown in BBL (their Fig.\,13) hints at the presence of jets 
interacting in several regions with the interstellar medium. Interactions between jets and dense molecular clouds 
produce strong shocks which have been shown to be possible causes of megamaser emission (e.g.\ Gallimore et al.\ 
\cite{gallimore01}; Peck et al.\ \cite{peck03}). Also the profile of the spectrum does not contradict a `jet-origin' 
of the detected maser emission in 3C\,403. If we assume a symmetric molecular distribution, the two observed features 
could be interpreted as the red-shifted and blue-shifted counterparts of the maser line, arising from opposite jets 
close to the core. 

Interestingly, the accretion disk scenario is a viable alternative, making use of the expected nuclear obscuring layer 
that should be oriented almost edge-on (see Chiaberge et al.\ \cite{chiaberge02}). As in the case of the Seyfert galaxy 
NGC~4258 (e.g.\ Miyoshi et al.\ \cite{miyoshi95}), the spectrum should then show three distinct groups of features: one 
centered at the systemic velocity of the galaxy (originating along the line of sight to the nucleus) and two groups 
symmetrically offset from the systemic velocity, arising from those parts of the disk that are viewed tangentially. 
If the latter are the two lines we are observing, the rotational velocity of the disk is $\sim$100\,\kms. In NGC~4258 
the systemic features are dominant, likely because the disk is slightly inclined, thus amplifying the southern radio jet. 
In 3C\,403 the systemic lines seem instead to be missing. To explain this fact by following the hypothesis proposed 
in Henkel et al.\ (\cite{henkel98}), we could argue that the circumnuclear disk is thin and not perfectly edge-on, 
thus not amplifying the radio flux of the core. In this case the background source would arise from the receding 
counter jet that may be too weak, because of relativistic dimming, to provide enough `seed' photons for a detection. 
So far, the orientation of the jets in 3C\,403 w.r.t.\ the plane of the sky has not been studied in detail.
Higher resolution maps are needed to determine which of the two scenarios, accretion `disk-maser' or `jet-maser'
emission, is most favorable to explain the observational data. 
\section{Conclusions}
Our discovery of water maser emission in the \frii\ galaxy 3C\,403 is:

$\bullet$ the first detection of a water megamaser in a radio-loud galaxy

$\bullet$ one of the very few direct detections of molecular gas, and the first indicating very high density 
($>$ 10$^{7}$\,cm$^{-3}$), in radio-loud galaxies

$\bullet$ another argument in support of the presence of molecular gas near the nuclear engine of \frii\ galaxies 

$\bullet$ a starting point to investigate, through follow-up high-resolution observations, the optically obscured inner 
parsecs of \frii s. Such observations will allow us to discriminate between nuclear `disk-' and `jet-maser' emission,
the latter inferring an interaction between the nuclear jet(s) and the ambient gas.
\begin{acknowledgements}
We are indebted to the operators at the 100-m telescope for their assistance with the observations. AT wishes 
to thank the MPIfR for its hospitality during the observing run. This research has made use of the NASA/IPAC 
Extragalactic Database (NED) which is operated by the Jet Propulsion Laboratory, California Institute of Technology, 
under contract with the National Aeronautics and Space Administration. 
\end{acknowledgements}

\end{document}